\documentclass{PoS}
\usepackage{lineno}

\title{Prospects of Earth-skimming neutrino detection with HAWC}

\ShortTitle{Prospects of Earth-skimming neutrino detection with HAWC}

\author{\speaker{H. Le\'on Vargas}\\
        Instituto de F\'isica, Universidad Nacional Aut\'onoma de M\'exico, M\'exico\\
        E-mail: \email{hleonvar@fisica.unam.mx}}

\author{for the HAWC Collaboration\footnote{for collaboration list see PoS(ICRC2019)1177 \newline and https://www.hawc-observatory.org/collaboration/icrc2019.php}}

\abstract{Searches for Earth-skimming neutrinos using volcanoes have not yet been achieved, but it is a promising technique for the detection of neutrinos above 1 PeV. The HAWC observatory is located in the vicinity of the highest mountain in Mexico, the Pico de Orizaba volcano, which shields the detector from quasi-horizontal very high-energy (VHE) muons. The large amount of shielding, up to 8 km of rock, enables the suppression of the large VHE muon background and makes neutrino detection possible. In this work we present the first steps towards the implementation of the Earth-skimming technique for the indirect measurement of tau neutrinos with HAWC. The results include the description of the charged lepton tracking reconstruction algorithm developed for this study and the initial analysis of the background of VHE muons using half a year of data.}

\FullConference{36th International Cosmic Ray Conference -ICRC2019-\\
		July 24th - August 1st, 2019\\
		Madison, WI, U.S.A.}

\begin{document}

\section{The HAWC Observatory}

The High-Altitude Water Cherenkov (HAWC) observatory was designed to be an instrument sensitive to Extensive Air Showers (EAS) in the TeV energy regime \cite{Crab}. It is located at approximately 4100 m a.s.l. in the Sierra Negra volcano, in the state of Puebla, Mexico. The main array of the observatory is composed of 300 cylindrical Water Cherenkov Detectors (WCDs), each having a diameter of 7.3 m and 4.5 m deep. The detector array is closely packed, covering a surface of approximately 22,000 m$^{2}$. Each WCD is filled with approximately 200,000 litres of purified water, and  instrumented with four upward facing photomultiplier tubes (PMTs) anchored at the base of the WCD. A 10-inch PMT is located at the center of the WCD, surrounded by three 8-inch PMTs that form an equilateral triangle around it. The sensitivity of HAWC to gamma-ray induced EAS, with energy above 1 TeV, is currently the best for a ground array. In this work we present the first results of an alternative use of HAWC, the indirect detection of ultra-high-energy neutrinos, that is based on the proximity of the observatory to the highest volcano in Mexico: the Pico de Orizaba, that reaches an altitude of 5 610 m a.s.l \cite{inegi}. The indirect search of neutrinos uses data acquired by the EAS trigger of HAWC, as it is described below.

\section{Earth-skimming neutrino detection with HAWC}

Early proposals of the Earth-skimming neutrino detection method using mountains go back to almost two decades, see for instance \cite{Neut-Fargion,Neut-Feng}. Recent experimental efforts include the limits established by the MAGIC \cite{Neut-MAGIC} and Pierre Auger \cite{Neut-PA} Collaborations. The current implementation of this detection method using the HAWC observatory was proposed in \cite{HAWC-Neut}, and it is based on the modular design of HAWC that allows to use the observatory as a very large particle tracker, considering each WCD as pixel for track reconstruction. To obtain good quality tracks, we can use the signal measured in each of the four PMTs installed in a WCD to define the quality of the pixels, i.e. by requiring that more than one PMT above a certain charge threshold is active in the event\footnote{By having more than one PMT hit per pixel, we can tell if the total amount of light detected is biased by a single PMT, as for instance caused by a vertical muon hitting it.}. The signal that we are interested on this analysis consists on a series of pixels being activated continuously, with propagation between them compatible with the speed of light.  

The most interesting flavour of neutrinos to be detected using the Earth-skimming method using volcanoes as targets are tau neutrinos. This is because the product of the neutrino-nucleon interaction in the target will produce a charged lepton of the same flavor as the incoming neutrino. Ultra-high-energy charged tau leptons are interesting due to their much lower radiative energy loss while propagating in matter, compared to muons, that allows them to be able to escape the mountain. A charged tau will decay quickly after production, causing an atmospheric shower with quasi-horizontal propagation. Due to the ultra-high-energy of the charged tau, the resulting shower will be highly collimated \cite{HAWC-Neut}. As a result of this, and due to the large dimensions of the HAWC WCDs, the resulting signal in HAWC would look like a track\footnote{This was tested using the standard full MC simulation of HAWC.}. This is the principle of detection of Earth-skimming neutrinos using the HAWC observatory.

\subsection{Triggering of candidate signals using the HAWC shower data}

The first step in the analysis consists on using the raw data obtained with the EAS trigger. The current HAWC trigger rate is of approximately 25 kHz and it is based on a simple multiplicity trigger that requires 28 PMTs with signal within a time window of 150 ns \cite{HAWC-NIM}. The basis that allows to use the same triggered data for this analysis as for the regular EAS search is that the event trigger window has a width of 1.5 $\mu$s. Even in the case of a very inclined shower, its propagation through the whole array takes approximately a few hundred nanoseconds, giving enough time in the triggered window to contain isolated tracks. In the case of small events (with less than 100 PMTs with signals in the event), the EAS trigger can actually be initiated by a single track, as it will be shown below. This is possible due to the high rate of vertical muons that, together with the presence of a short track-like signal that hits from 8 to 12 PMTs (in four WCDs, with two to three PMT hits per WCD), is enough to trigger the HAWC DAQ. For details about the HAWC DAQ system please refer to \cite{HAWC-NIM}. The trigger for track-like signals works following these steps:
\vspace*{-0.15cm}  
\begin{enumerate}
\item The raw data is reconstructed using the standard HAWC tools. The reconstruction software calibrates the PMT signals (for both charge and time). The shower plane is reconstructed among other EAS observables, but we skip certain functionalities such as shower energy estimation in order to speed up the data processing. The shower plane reconstruction is useful to veto the PMT signals associated with it. The PMT hits associated to the shower plane are not used in the search for track-like signals.
\vspace*{-0.10cm}  
\item The PMT signals from the event are time ordered and then processed by the track trigger. The algorithm starts with the first PMT with a signal in the shower event, and a second active PMT is looked for in a different WCD, within a time window given by the maximum distance between tanks in the HAWC array (44.8 m) and the speed of light. The neighbouring tanks are defined by a look up table that allows to search for propagation not only to the closest set of WCDs, but also to the following set (in the radial direction, starting from the first WCD associated to the track candidate), in order to avoid issues such as track splitting due to inactive WCDs that could artificially stop the propagation of the signal. A WCD is defined as active, in this analysis, when it contains at least one PMT with a charge measurement above 4 PEs\footnote{ Several possible charge thresholds were studied, and a good compromise between reducing noise and disk space usage resulted in a threshold value of 4 PEs per PMT.}. 
\vspace*{-0.10cm}  
\item The track event window is defined by the maximum amount of hits that are continuously propagated between neighbouring WCDs, as long as the signals of subsequent pair of hits are within the allowed time window  described above (150 ns). When a pair of time-ordered neighbouring hits is separated by more than the allowed time window, or the signals are separated beyond two adjoining WCDs, the track triggered event is defined. More than one track triggered event is allowed in the same shower event.
\vspace*{-0.10cm}  
\item The track triggered event is stored if it comprises more than nine PMT hits. For very bright tracks that would hit all four PMTs inside a WCD, this implies a minimum requirement of three WCDs being hit in order to store the track triggered event for further processing. For each track triggered event, all the information from the PMTs that are hit within the time window of the track event are stored as per pixel information, even from those that in principle are not related to the track candidate.  The pixel information is the following: total amount of charge deposited in each WCD, average arrival time of the hits and the number of hits within each pixel. Only the central PMT information is stored for each pixel, in order to study afterwards if the pixel information is not biased by the signal of the PMT with the larger surface and quantum efficiency.
\end{enumerate}
\vspace*{-0.10cm}  
The track triggered data contains a large number of candidate signals. However, most of these are primarily due to small inclined showers that failed to be identified by the EAS reconstruction algorithms. In order to maintain a reasonable number of triggers, a condition is set on the candidate shower events. If they contain more than 100 active PMTs they are discarded. This value was also optimized to find a balance between rate of signals and disk space usage. The candidate signals found by the track trigger (approximately with a rate of 16.7 kHz), are passed to a second code that performs the track reconstruction, using as input the pixels selected by the track trigger algorithm.

\subsection{Tracking algorithm}

The algorithm used to find track candidates is very simple, it searches for straight lines propagating through the HAWC main array. We decided that it was not necessary to use a more sophisticated algorithm since the number of pixels involved is restricted by the number of active WCDs in the detector, which for this analysis we set to at most 100 WCDs. At this stage the actual position of the different WCDs in the detector array is used to look for propagation through pixels at the speed of light, allowing a time window of $\pm$ 40 ns for the propagation between two pixels to be considered valid. The time window value was optimized using Monte Carlo simulations of muons thrown at the HAWC array from different orientations through the detector. The tracking proceeds using the time ordered pixel signals. At this point a requirement of having at least two PMT hits per pixel is included. The track searching of the algorithm over the triggered data is iterative, i.e. after a track is identified, the pixels involved in the track are removed from the track triggered event and the track finding is run again over the remaining pixels, until all possible seeds for the track finder are tested.

It is at this stage that the angular properties of the reconstructed track are calculated, and only the pixel information of the WCDs involved in the track is saved for further processing. The tracking algorithm reconstructs the trajectories by using straight lines that connect the centers (in the X-Y plane, using the HAWC local coordinate system) of the pixels with signals consistent with propagation at the speed of light among them. The first two pixels with signals define an initial direction to search for the propagation trajectory, and subsequent participating pixels are allowed to change the original direction by up to 30 degrees, a value optimized using MC simulations. Tracks that comprise at least three pixels are stored as track candidates. The elevation and azimuth properties of the track are defined based on the locations and dimensions of the WCDs involved in the track. Since the main array of HAWC is located on a flat surface, this limits the elevation angles that are possible to study. However, one can notice on the preliminary results presented on this work that the accessible solid angle covers the most interesting region of the Pico de Orizaba, the one with the largest overburden. The overall information of the original shower event (number of  WCDs with PMT hits, active number of pixels in the track trigger event and trigger time) is stored along with the track candidate information (pixel by pixel), in order to make the selection cuts that are discussed below. For the reported average charge deposited in each pixel by a track, we scaled the charge measured by the central PMT to make it compatible with the charge measured by the peripheral PMTs located in the same WCD \cite{Crab}. We use the average charge deposited in the pixels of the track as a proxy for the energy of the event. A set of Monte Carlo simulations, produced using the standard tools of HAWC, were prepared in order to evaluate the tracking algorithm. The difference with respect to the EAS simulations is that in this case we injected single particles, positive muons in the energy range [100 GeV, 100 TeV], instead of air showers. The performance of the tracking algorithm is discussed in the next subsection. The rate of track candidates in data is of approximately 420 Hz.

\subsection{Filtering of candidate tracks}

\begin{figure}[h]
\centering
\begin{minipage}{.45\linewidth}
  \includegraphics[width=\linewidth]{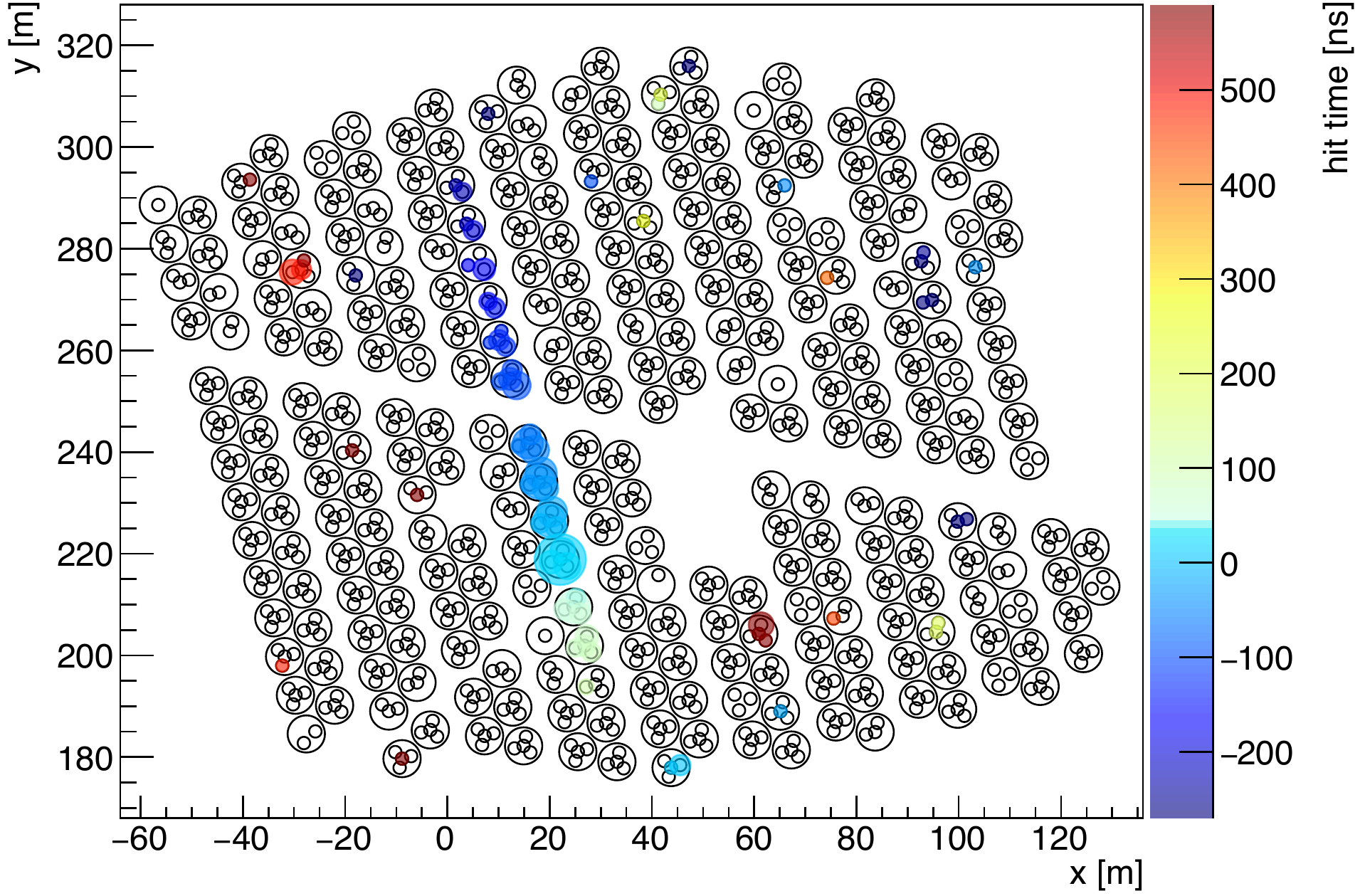}
  \caption{Display of the signals produced by an event acquired with the EAS trigger. The event  was afterwards identified as containing a track by our track finding algorithms. See the text for details.} 
  \label{fig-1}
\end{minipage}
\hspace{.05\linewidth}
\begin{minipage}{.45\linewidth}
  \includegraphics[width=\linewidth]{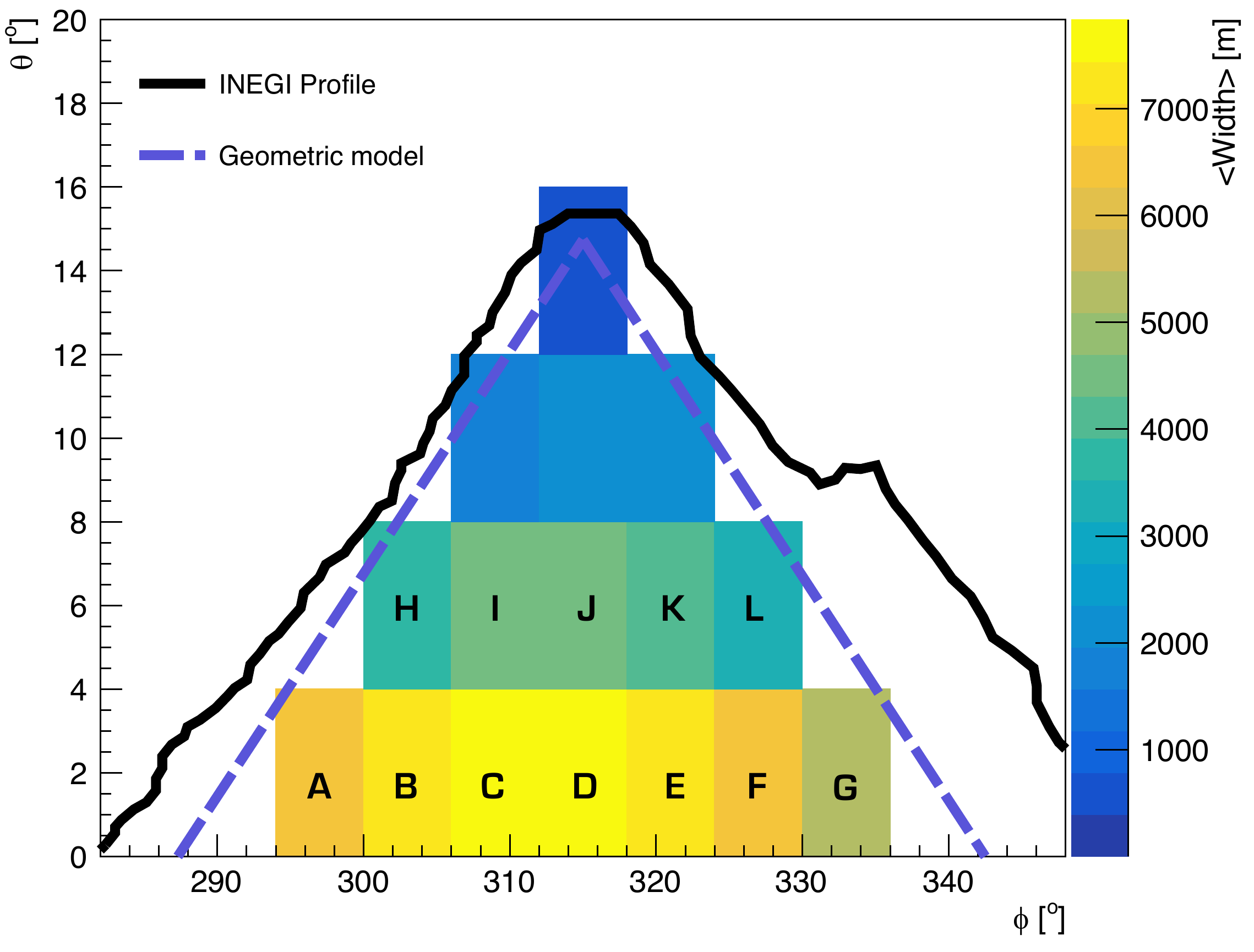}
  \caption{Definition of the cells used in the analysis of the average track charge as a function of the mean value of overburden for a given section of the Pico de Orizaba volcano.}
  \label{fig-2}
\end{minipage}
\end{figure}

Despite the selection requirements used in the tracking algorithm, the track candidate sample is contaminated with very inclined small showers (showers that hit less than 30 WCDs). In order to get rid of this contamination, a filtering process is applied to all track candidates to obtain a sample of isolated tracks. The cuts are illustrated using the figure \ref{fig-1}, which shows an event display of an EAS triggered event that actually contains a track. The color code indicates the time of each PMT hit within the trigger window and the size of the circles with color is proportional to the measured charge.
Table \ref{tab-1} shows the properties from the full shower event and the reconstructed track. All the cut values used on this analysis were selected based on studying a large number of track candidates in an event display. The filtering of the tracks is based on the following variables:
\vspace*{-0.15cm}  
\begin{itemize}
\item Low Charge (LC) activity: This variable compares the total activity in the shower event (number of active WCDs in the event or N$\mathrm{_{WCDs}}$) with the length of the track candidate, quantified by the number of pixels that make up the track (TL). The name comes from the fact that a single PMT with a signal within a WCD sets it as active in the standard HAWC analysis, without any charge threshold for the PMTs. Thus, the variable is simply given by LC = N$\mathrm{_{WCDs}}$/TL. For the current analysis the cut on this variable is that the event should have an LC $<$ 6.
\vspace*{-0.2cm}  
\item High Charge (HC) activity: This variable compares the total number of active pixels in the track event (NPixel$_{\mathrm{A}}$) to those that belong to the track candidate. This variable is defined as the ratio HC = NPixel$_{\mathrm{A}}$/TL. For this analysis we require HC $\leq$ 1.5.
\vspace*{-0.2cm}  
\item Track quality: We verify that the track candidate can be fit using a linear function that includes at least 75\% of the pixels identified as part of the track. We do not enforce that 100\% of the pixels can be fit, since we expect that some of the highly collimated signals arriving to the detector may hit more than just one line of WCDs. A good example of this is shown in figure \ref{fig:cellEev}. We also check that the propagation of the signals between the first and last pixel of the track is consistent with the speed of light.
\vspace*{-0.2cm}  
\item Minimum length: This selection criteria was established based on an analysis of possible track contamination due to combinatorial background. For the full six month data sample, we randomized the PMT locations while keeping the charge and timing information of the original shower event, and then run our track finding algorithms. We found a small number of fake tracks, but the maximum length of these was of three pixels. We require that all tracks have a TL $\geq$ 4 pixels.
\end{itemize}

\begin{table}[!htbp]
\centering
\begin{tabular}{ccc|cccccc}	
\hline
\multicolumn{3}{c}{Event characteristics} & \multicolumn{6}{c}{Track characteristics}  \\
\hline
PMTs & N$\mathrm{_{WCDs}}$ & NPixel$_{\mathrm{A}}$ & TL   & LC   & HC & $\mathrm{<Track_{Charge}>}$ & $\phi$ [$^{\circ}$] & $\theta$ [$^{\circ}$] \\
\hline
96 & 42 & 13 & 10   &  4.2 & 1.3 &  55.4 PEs & 254.6 $\pm$ 2.8 & 1.0 $\pm$ 1.9 \\
\hline
\end{tabular}
\caption{Properties of the original EAS triggered event, shown in figure \ref{fig-1}, and those of the reconstructed track contained on it.}
\label{tab-1}
\end{table}

The average angular aperture between an injected muon in the Monte Carlo and its reconstructed track is of approximately 3.0$^{\circ}$ in the energy range from 10 GeV to 100 TeV, for the angular region used on this study, shown in figure \ref{fig-2}. The rate of tracks that pass all the cuts is drastically reduced to 0.5 mHz. The high quality of these tracks was tested using an event display, that allowed to confirm that the cuts described above efficiently remove small showers misidentified as tracks.

\section{Preliminary results}

\begin{figure}[h]
\centering
\begin{minipage}{.30\linewidth}
  \includegraphics[width=\linewidth]{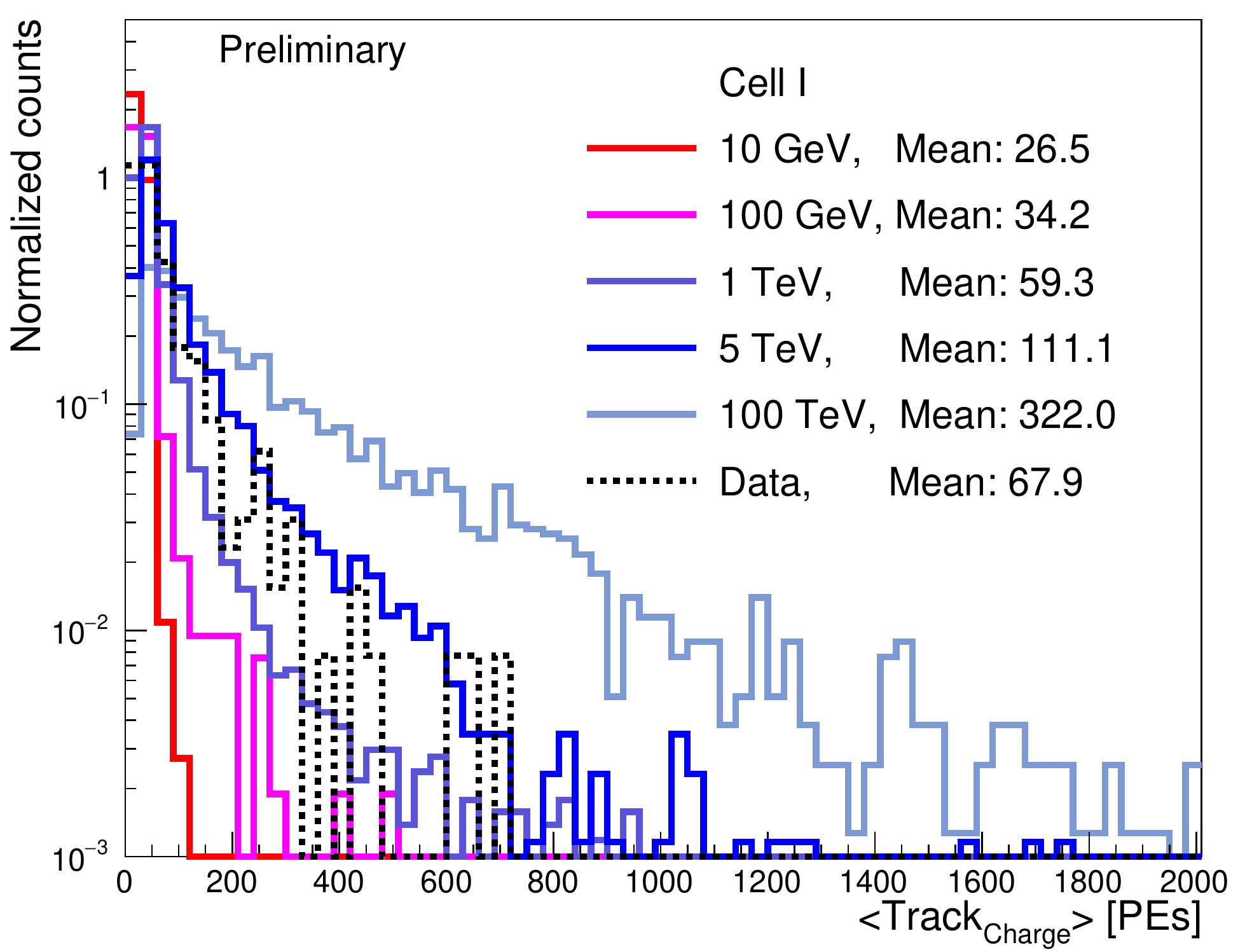}
  \caption{Average charge deposit for tracks in the cell I.}
  \label{fig:CellI}
\end{minipage}
\hspace{.03\linewidth}
\begin{minipage}{.30\linewidth}
  \includegraphics[width=\linewidth]{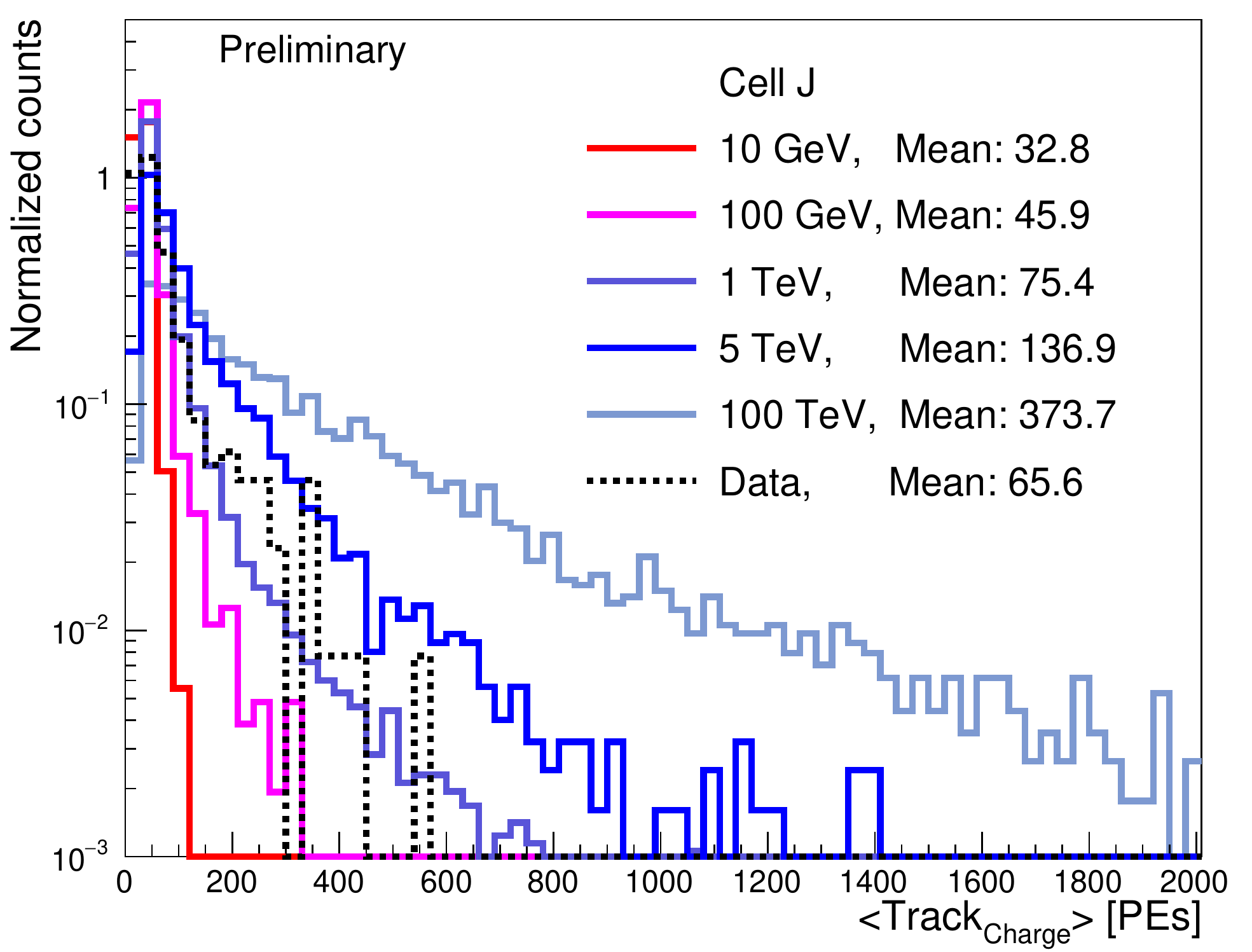}
  \caption{Average charge deposit for tracks in the cell J.}
  \label{fig:CellJ}
\end{minipage}
\hspace{.03\linewidth}
\begin{minipage}{.30\linewidth}
  \includegraphics[width=\linewidth]{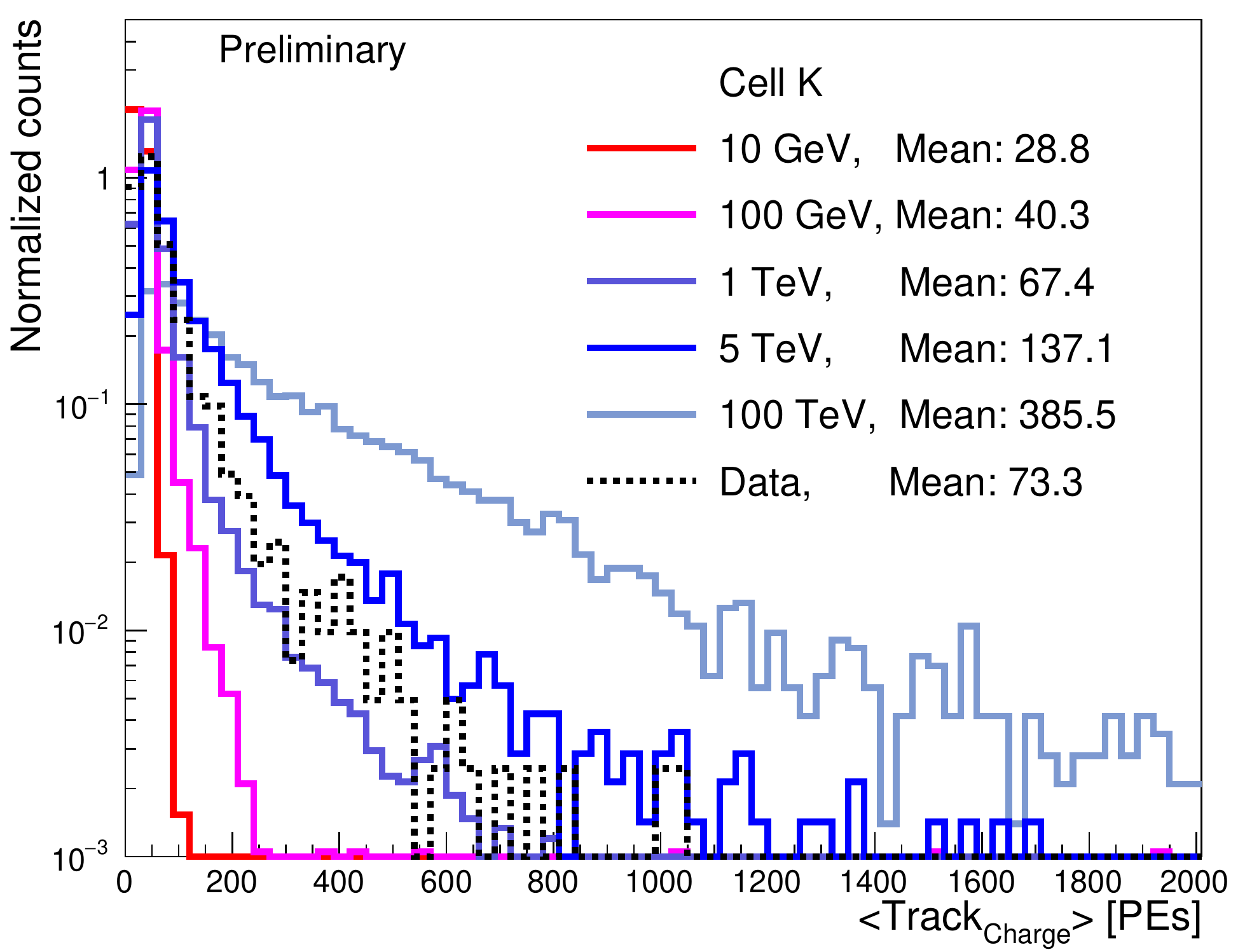}
  \caption{Average charge deposit for tracks in the cell K.}
  \label{fig:CellK}
\end{minipage}
\end{figure}

\begin{figure}[h]
\centering
\begin{minipage}{.30\linewidth}
  \includegraphics[width=\linewidth]{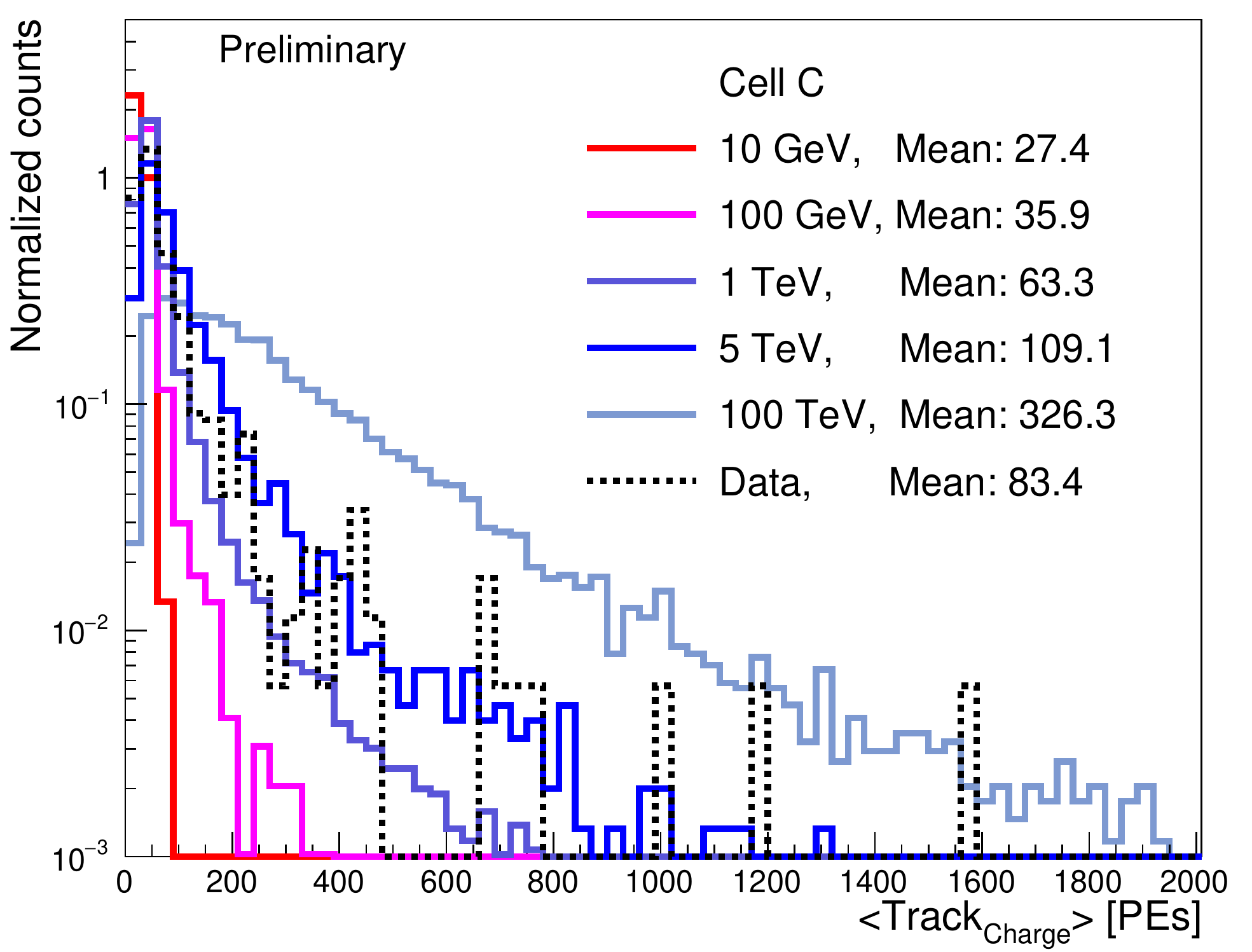}
  \caption{Average charge deposit for tracks in the cell C.}
  \label{fig:CellC}
\end{minipage}
\hspace{.03\linewidth}
\begin{minipage}{.30\linewidth}
  \includegraphics[width=\linewidth]{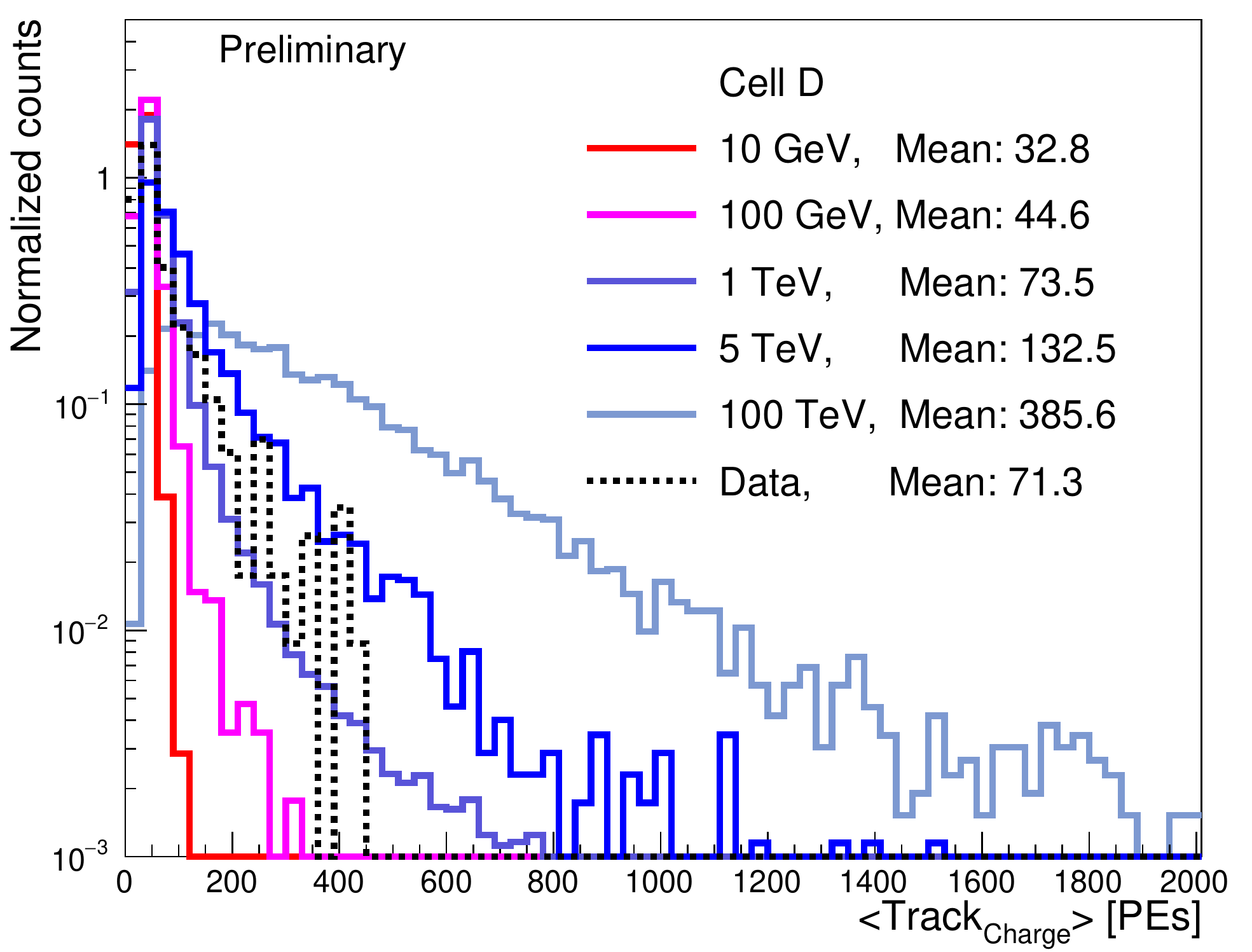}
  \caption{Average charge deposit for tracks in the cell D.}
  \label{fig:CellD}
\end{minipage}
\hspace{.03\linewidth}
\begin{minipage}{.30\linewidth}
  \includegraphics[width=\linewidth]{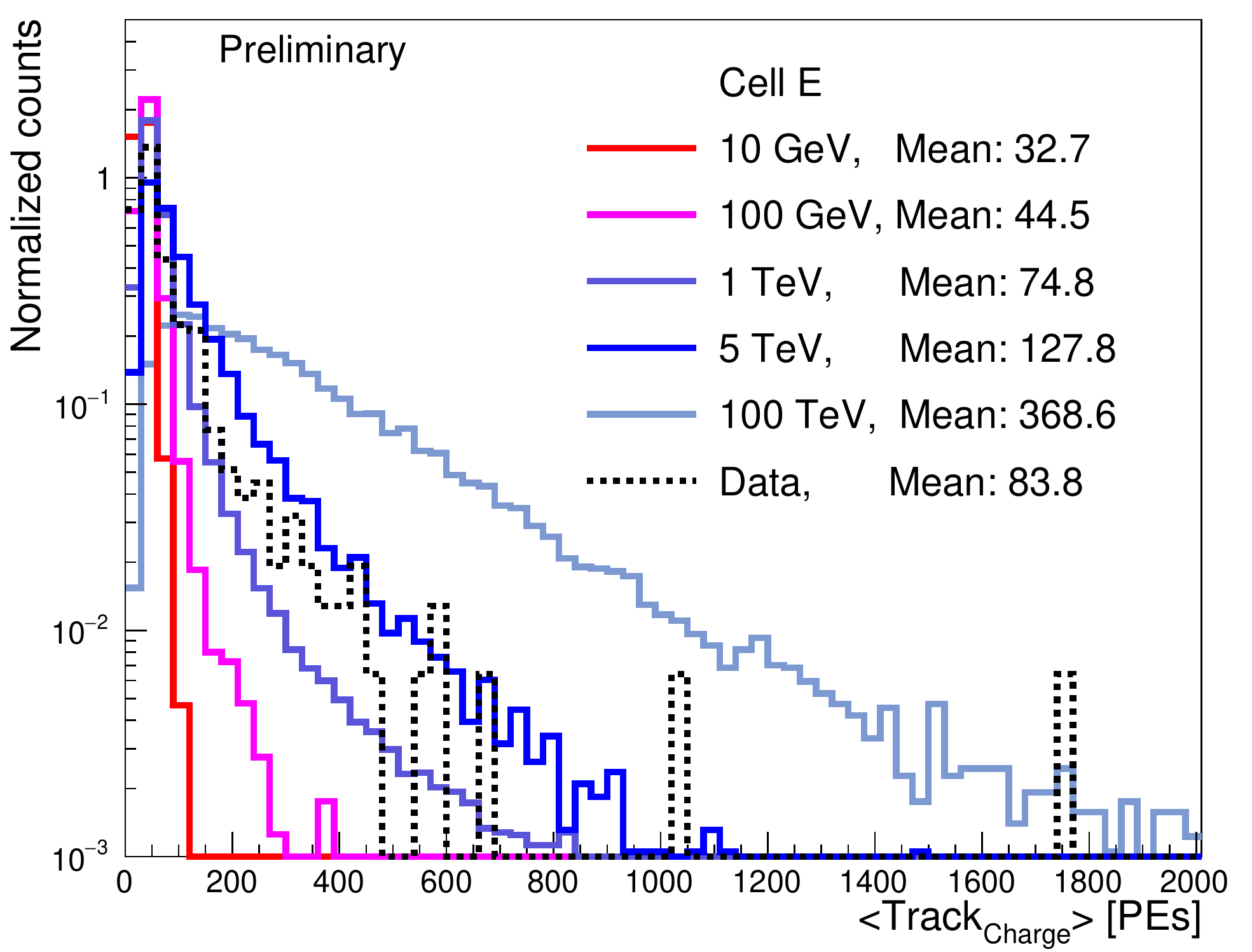}
  \caption{Average charge deposit for tracks in the cell E.}
  \label{fig:CellE}
\end{minipage}
\end{figure}

\begin{figure}[h]
\centering
\begin{minipage}{.45\linewidth}
  \includegraphics[width=\linewidth]{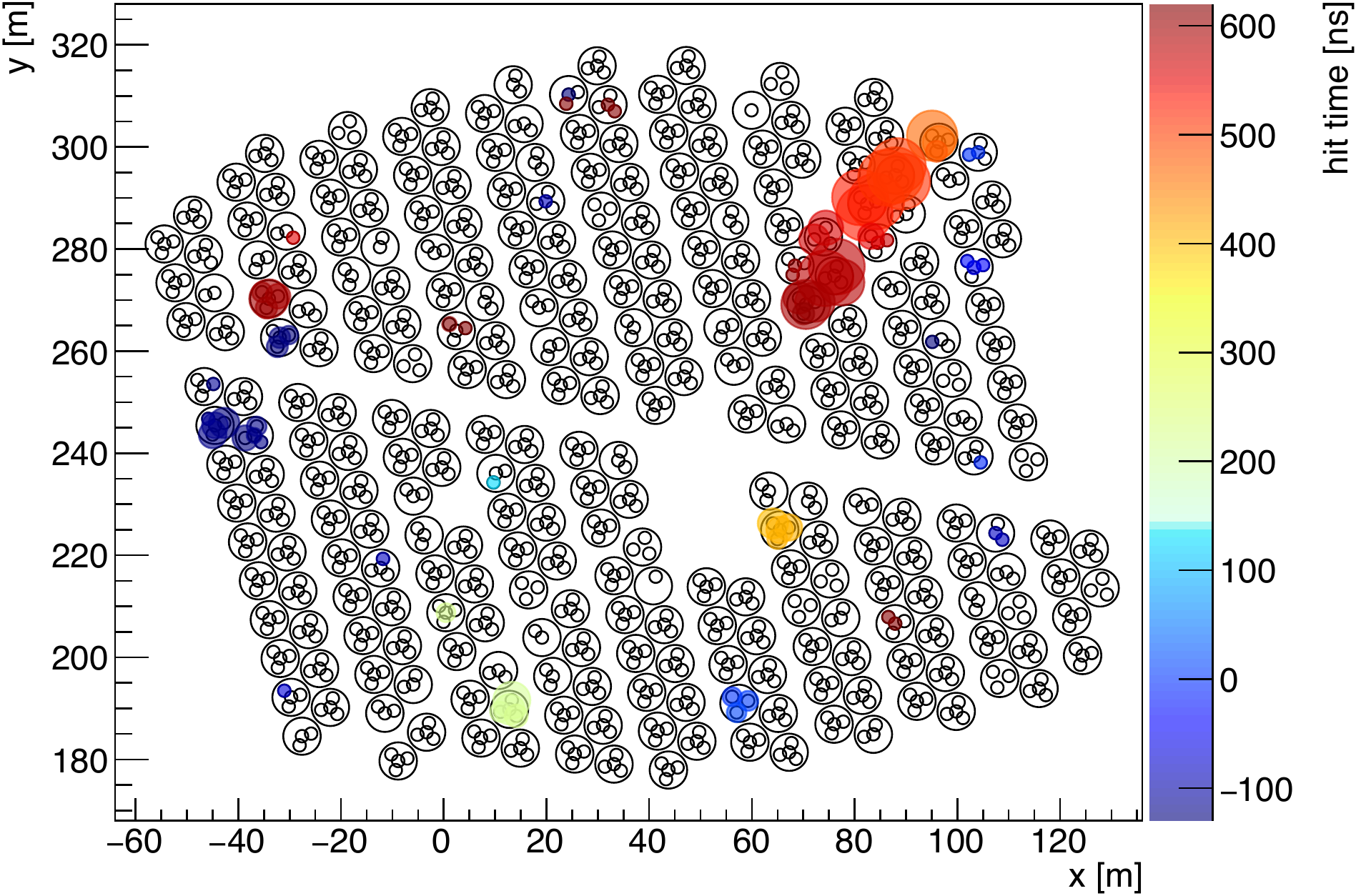}
  \caption{Event display of a track-like signal whose reconstructed direction is consistent with the region with largest overburden in the Pico de Orizaba volcano (cell C in figure \ref{fig-2}).}
  \label{fig:cellCev}
\end{minipage}
\hspace{.05\linewidth}
\begin{minipage}{.45\linewidth}
  \includegraphics[width=\linewidth]{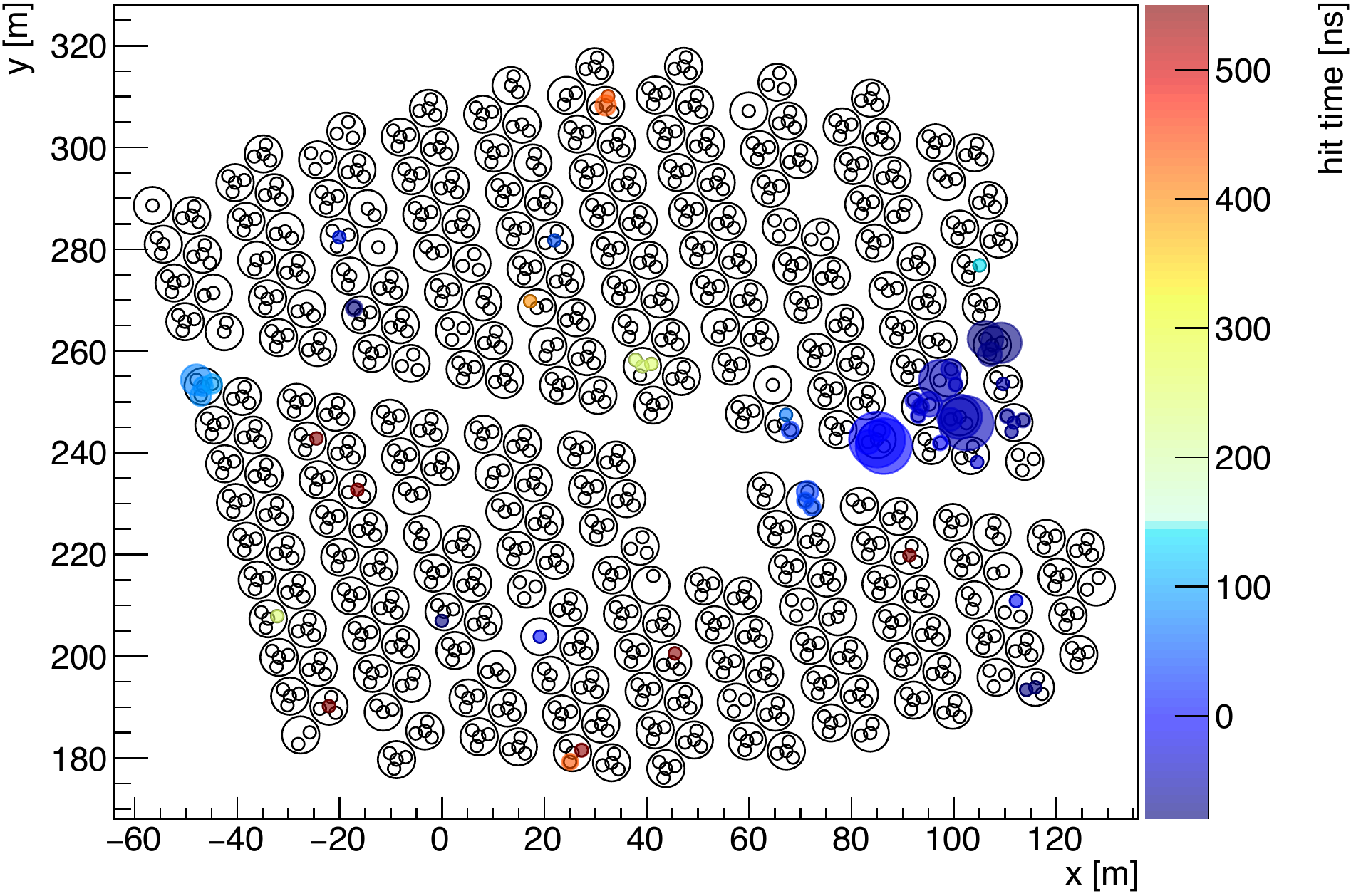}
  \caption{Event display of the track-like signal with the largest average charge deposit in the whole data sample and whose reconstructed direction is consistent with the cell E in figure \ref{fig-2}.}
  \label{fig:cellEev}
\end{minipage}
\end{figure}

The following results were obtained using approximately 181 days of data from the EAS trigger. For this analysis, we divided the solid angle covered by the Pico de Orizaba volcano in several rectangular bins of six degrees in azimuth and four degrees in elevation. Figure \ref{fig-2} shows, with a solid black line, the profile of the Pico de Orizaba volcano as seen from the center of the HAWC main array, using data from the Mexican National Geography and Statistics Administration (INEGI) \cite{inegi}. The purple dashed lines show the geometrical approximation used for the effective area calculation in \cite{HAWC-Neut}. The rectangular bins show the sections of solid angle used to study the properties of the reconstructed tracks as a function of the average width of the volcano that a track pointing back to that solid angle have to pass through. The average width of the volcano in each direction, shown with the color code in figure \ref{fig-2}, was calculated using INEGI data. Figures \ref{fig:CellI}, \ref{fig:CellJ} and \ref{fig:CellK} show the distributions of the average charge, deposited in each pixel, of the tracks whose trajectory comes from the cells I, J and K defined in figure \ref{fig-2}. The distributions are normalized so that the area under each distribution is equal to 100, so they basically describe the probability that a track from each data sample would deposit a given amount of charge in the HAWC array. In each of the plots, the real data distribution (shown with a black dotted line) is compared to the charge distributions obtained with samples of mono-energetic positive muons thrown towards HAWC from the angular regions defined by each cell. The results shown in figures \ref{fig:CellI}, \ref{fig:CellJ} and \ref{fig:CellK} correspond to the intermediate overburden region of the volcano, as one can see from figure \ref{fig-2}. One can notice that the charge deposits from real data are never above the expected value for 5 TeV muons. On the other hand, figures \ref{fig:CellC}, \ref{fig:CellD} and \ref{fig:CellE} show the charge distributions from tracks with directions pointing back towards the base of the volcano, i.e. the region with the largest overburden. In figure \ref{fig:CellC} one can notice that there is a track whose deposited charge is well beyond the expected range for muons with energies of 5 TeV. The average charge deposited in each pixel from that track is of 1561.7 PEs. Another interesting case is presented in figure \ref{fig:CellE}, where there is a track with an average charge deposit per pixel of 1744.8 PEs. The event displays of these two track-like signals are shown in figures \ref{fig:cellCev} and \ref{fig:cellEev} respectively.

\vspace*{-0.3cm}  

\section{Summary}

We presented preliminary results for the search of Earth-skimming neutrinos with the HAWC observatory, using six months of data. We identified two track-like signals coming from the base of the volcano (> 18 km.w.e.) that produce very large charge deposits in the detector array. Further investigation of these events is in progress.

\acknowledgments
This study is dedicated to the memory of Elia Judith Vargas Cisneros. This work was supported by the Consejo Nacional de Ciencia y Tecnolog\'ia, M\'exico, grant 254964.
For a complete list of acknowledgements see: https://www.hawc-observatory.org/collaboration/icrc2019.php.

\end{document}